# Graphoepitaxy for Pattern Multiplication of Nanoparticle Monolayers

Mark E. Ferraro, Roger T. Bonnecaze, and Thomas M. Truskett

McKetta Department of Chemical Engineering and Texas Materials Institute, The University of Texas at Austin, Austin, TX USA 78712

*23 October 2014*

We compute the free energy minimizing structures of particle monolayers in the presence of enthalpic barriers of a finite height $\beta V_{ext}$ using classical density functional theory and Monte Carlo simulations. We show that a periodic square template with dimensions up to at least ten times the particle diameter disrupts the formation of the entropically favored hexagonally close-packed 2D lattice in favor of a square lattice. The results illustrate how graphoepitaxy can successfully order nanoparticulate films into desired patterns many times smaller than those of the prepatterned template.

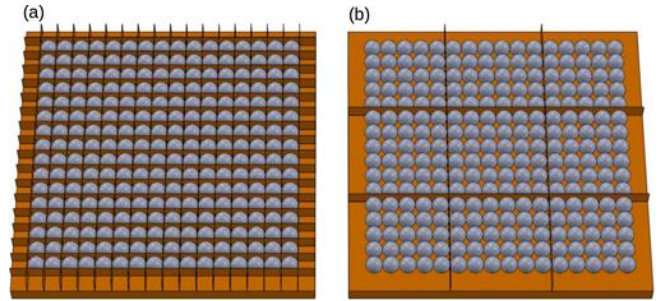

Fig. 1: Monolayers assembled via deposition onto a substrate with repulsive barrier templates separated by a distance commensurate with the square lattice unit cell. The template in (a) trivially fixes each particle in its ideal location. The template in (b) uses sparser templating to direct assembly into the same structure.

Assembly of particles into targeted structures can play an important role in the manufacture of novel materials and devices. Particles with specific compositions and shapes can be synthesized [1], solution processed, and cast into micro- or nanostructures with desired optical, electronic, physical or chemical properties [2,3]. Recent work has shown that the symmetry of the assembled structure can be tuned by particle shape [4,5], polydispersity [6], anisotropic interactions [7,8], and the attachment of chemically active ligands [9,10]. It is even possible to use inverse methods of statistical mechanics to discover new forms of pairwise interactions that stabilize a desired structure [11-14]. In practice, particles often need to be directed externally to form the required lattices [15-17]. The role of an external field in the assembly of particulate systems, however, is still incompletely understood.

Here, we consider a model for the solution-based deposition and assembly of spherical particles onto a smooth, attractive surface with regularly spaced enthalpic barriers that represent, for example, prepatterned chemical or topographical features. Particles from a contacting fluid suspension can adsorb onto or desorb from the substrate, where they diffuse in the presence of the periodic barriers. As the liquid film dries, the equilibrium surface concentration increases until only a monolayer of deposited particles remains. For hard-sphere particles on a pattern-free substrate, a disordered liquid-like structure is favored for areal packing fractions $\eta < 0.72$. Above $\eta = 0.72$, a hexagonally close-packed (hcp) solid forms [18]. To successfully direct the particles into an alternative packing at high η, an appropriately designed external field provided by the substrate must favor the targeted structure over the otherwise entropically favored hcp lattice. This type of graphoepitaxial assembly has been successfully used in other contexts, e.g., to direct the assembly of block copolymer films into targeted morphologies [19-28].

We study a model for such an elementary pattern multiplication process for deposited particulate films. We show that a smooth substrate with a periodic square pattern of enthalpic barriers is able to disrupt formation of the hcp lattice in favor of a square lattice structure as the surface concentration increases. Interestingly, this approach remains effective even when the barriers are separated by length scales significantly (~10x) larger than the particle diameter (i.e., the lattice constant), illustrating the promise of using graphoepitaxy for directing particulate assemblies of patterned monolayers.

In our model, 3D hard spheres—once adsorbed on the 2D substrate—interact with each other as exclusion disks in the grand canonical ensemble with an external field (representing the prepatterned barriers) and with a chemical potential set by the reservoir of particles in the contacting fluid suspension. Slow evaporation of the solvent concentrates the reservoir, thus raising the chemical potential and increasing the particle concentration on the surface. Increasing the particle-substrate attraction at constant reservoir concentration would have the same net effect. We model the thermodynamics and structure of the adsorbed monolayer as a function of chemical potential and template features (i.e., barrier height and separation) using classical density functional theory and Grand Canonical Monte Carlo (GCMC) simulations.

For our density functional theory predictions, we use Rosenfeld's Fundamental Measure Theory (FMT) [29], which provides accurate equilibrium density profiles for inhomogeneous hard-particle fluids with greater computational efficiency than Monte Carlo simulations [30]. The FMT framework is also ideal for answering inverse design questions for targeted self-assembly, which—while not addressed specifically here—is a fruitful area for future study. We adopt a version of FMT recently introduced by Roth et al. [31] which faithfully reproduces the structure of high-density fluid states as well as the first-order fluid-to-hcp phase transition. For computational efficiency, equilibrium FMT computations reported here use a matrix-free Newton iterative method developed by Sears and Frink [32]. For our 2D GCMC simulations of the same model system, we adopted a simulated tempering algorithm designed to study high-density hard-particle systems in the vicinity of the melting transition [33].

For our target structure, we choose a periodic square lattice, which is of interest for bit-patterned magnetic media [34]. The square lattice has a lower coordination number (4 vs. 6) and a less efficient packing structure than the hcp crystal, which means that, for hard spheres, graphoepitaxy will be required to make it the thermodynamically stable structure. The trivial graphoepitaxial solution would be to use substrates with parallel barriers that intersect to form adjacent square "boxes" with side lengths equal to twice the particle radius, $2R$, so that—once adsorbed—each particle is precisely confined to its desired lattice position (Figure 1a). This, of course, in addition to having other practical limitations, requires accurate pre-patterning on the same length scale as the target lattice, which negates the most powerful pattern amplification benefits of directed self-assembly. Thus, we consider here square templates of barriers separated by integer multiples of the particle diameter; that is, sparser patterns commensurate with the desired square lattice periodicity (Figure 1b). The barriers are energy step functions of width $2R$ and normalized "height" $\beta V_{ext}$ ($\beta=1/k_BT$, where $k_B$ is the Boltzmann constant and $T$ is temperature) acting on particle centers; i.e., they penalize any "overlap" of a particle with the infinitely thin line centered on the square template boundary. This template allows us to direct assembly of particles into periodic, close-packed square-lattice structures. We present results in terms of the difference in chemical potential, $\beta(\mu-\mu_{hcp})$, between the system of interest and the value where the 2D hcp solid becomes the thermodynamically stable adsorbed phase on a patternless substrate.

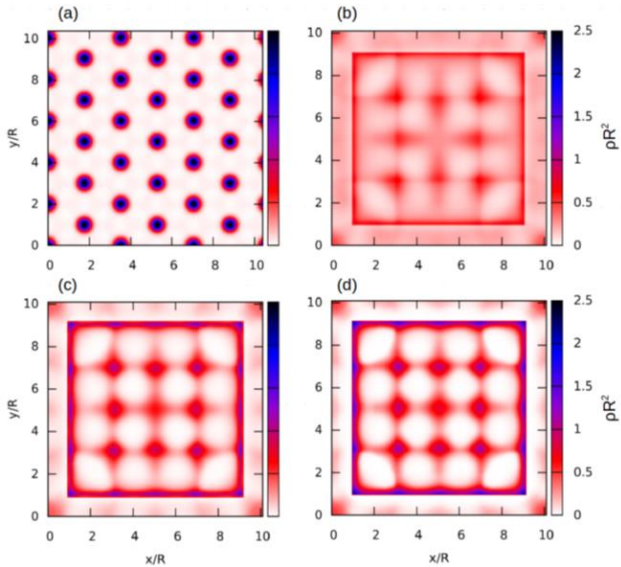

Fig. 2: Equilibrium density profiles with $\beta(\mu-\mu_{hcp}) = 2.40$ predicted using FMT. The external field is as illustrated in Fig 1(b), with barriers of height $\beta V_{ext}$ = (a) 0, (b) 2, (c) 5, and (d) 10, each separated by a distance of $10R$.

Fig. 2 shows the equilibrium density profiles predicted by FMT as a function of increasing barrier height of the template. In this figure the chemical potential [$\beta(\mu-\mu_{hcp})$ = 2.40] is higher than what would be otherwise necessary to form the hcp structure with a pattern-free substrate (see Fig. 2a). When particles absorb in the presence of the template, the equilibrium density profile qualitatively changes and is characterized by structures with square symmetry (Figs. 2b,c), even for a barrier height as low as $\beta V_{ext,}$= 2. For increasing values of barrier height, the density profiles become sharper and the square lattice structure more pronounced. It is interesting to note that the template feature separation in Fig 2 is nearly commensurate for both the hcp and square lattices, and yet the square symmetry is still clearly preferred in the presence of finite barriers.

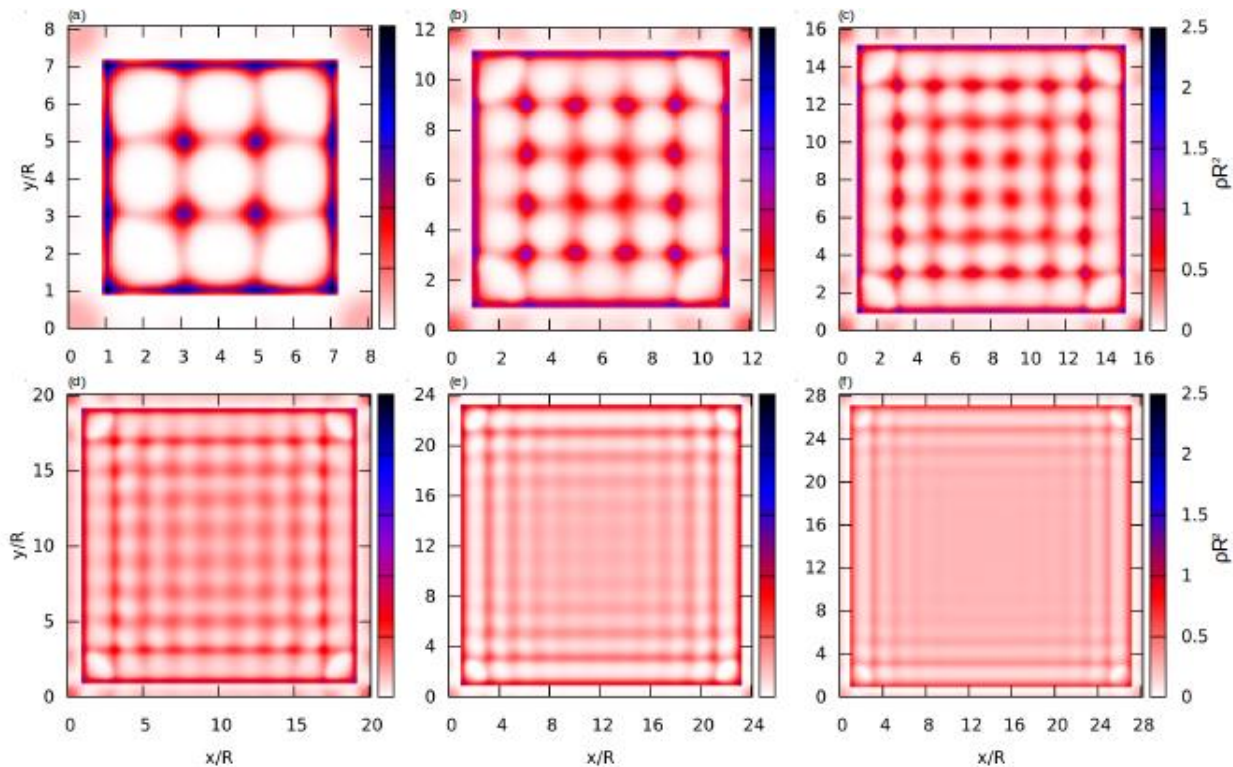

Fig. 3: Equilibrium density profiles predicted by FMT as a function of template feature spacing for $\beta(\mu-\mu_{hcp}) = 2.40$. Enthalpic barriers of height $\beta V_{ext} = 10$ are placed at each border of the displayed region. Template feature separation for each profile is (a) 8R, (b) 12R, (c) 16R, (d) 20R, (e) 24R, and (f) 28R.

The effect of template spacing from 8R to 28R for $\beta V_{ext} = 10$ and $\beta(\mu-\mu_{hcp}) = 2.40$ is shown in Fig. 3. Note that while the structuring weakens as the template spacing grows, the square symmetric order persists even when the spacing is an order of magnitude larger than the particle diameter. However, for template feature separations greater than 20R, the density peaks lower and spread and local maxima are not significantly greater than the bulk density. At a template spacing of 24R, the density signature of a square pattern in the center has practically vanished, which indicates the barriers are too far apart to promote long range ordering (see the Supplemental Material). As the wall separation increases to 28R, the density profile in the middle of the system becomes nearly flat.

While the FMT density profiles provide useful average information about the structural consequences of the template features, we turn to grand canonical Monte Carlo (GCMC) simulations to characterize the defect statistics from the underlying microstates and understand how they evolve with increasing chemical potential and surface concentration. The occupancy probabilities for a 10Rx10R square obtained from GCMC simulations of a periodically-replicated 9-square system is shown in Fig. 4. For chemical potentials of $\beta(\mu-\mu_{hcp}) = 1.87$, some of the square regions in the template are perfectly filled and others are missing one or more particles (Fig. 4b). Perfectly filled regions form square lattices. Regions with one or two missing particle form structures with point defects and those with two or three missing particles resemble 2D liquid-like structures. Furthermore, these defective structures explain the qualitative features observed in the FMT density profiles. These profiles are obtained in the grand canonical ensemble, meaning they comprise microstates with different numbers of particles. Microstates with vacancies allow rows and columns to freely translate, which contributes to the average density on the grid connecting square lattice coordinates. As $\beta\mu$ increases, the fraction of microstates containing the ideal 25 particles smoothly increases. Virtually zero defectivity is achieved for $\beta(\mu-\mu_{hcp}) = 5.87$ and $\beta V_{ext} = 10$ with the structural configuration shown in Fig. 4c.

The manner in which the defects in the square lattice naturally fill in as a function of chemical potential can be understood from a basic statistical mechanical argument. For inhomogeneous hard-disk systems, it can be shown that the position-dependent insertion probability is given by $P_{ins}(x,y) = \rho(x,y)\lambda^2/\exp(\beta\mu)$ [35] where $\lambda$ is the thermal de Broglie wavelength. The probability of successfully inserting a particle is proportional to the equilibrium number density $\rho(x,y)$ at that location. From the FMT density profiles in Figs. 2 and 3, we see that graphoepitaxy naturally places peaks in the density profile at the preferred square lattice sites. Thus, particles preferentially adsorb at point defects in the square lattice. As solvent evaporation occurs, the increasing chemical potential pushes more particles into these open lattice sites, driving the system toward a defect-free square lattice structure as shown in Fig. 4c.

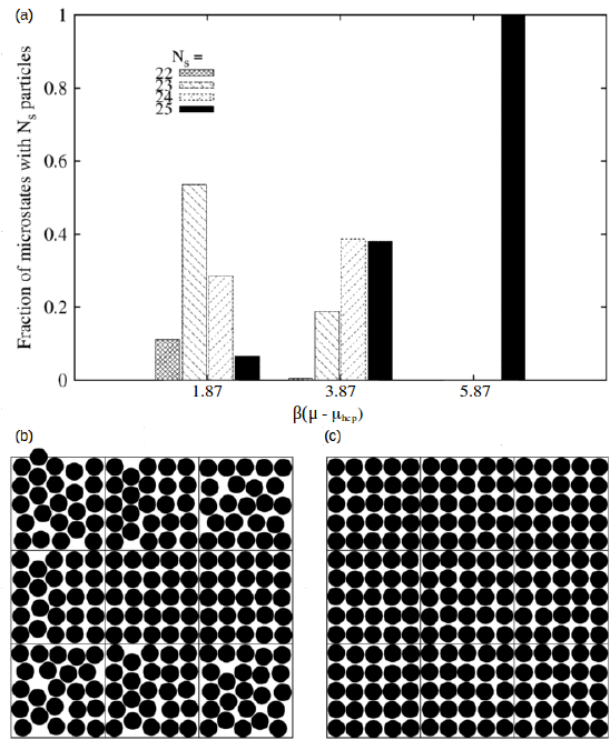

Fig. 4: (a) Fraction of microstates with $N_s$ particles as a function of chemical potential $\beta(\mu-\mu_{hcp})$. Here, a microstate is defined by a single periodic 10Rx10R square bounded by the external field from Fig. 2(d). Configurations shown in (b) and (c) are snapshots taken from the GCMC simulations where $\beta(\mu-\mu_{hcp}) = 1.87$ and 5.87, respectively. Lines in plots (b) and (c) represent thin barriers of height $\beta V_{ext} = 10$.

It should be noted that the wall separation need not be an exact integer multiple of the particle diameter. Square order will still be observed for integer separations plus an additional length that can be up to approximately half a particle radius. This is consistent with known closest packing results, which have shown that certain square packings require a significant expansion of the confinement area before an additional particle can be accommodated [36]. With our system of finite, fixed chemical potential and barrier height (unlike the maximum packing analogy), entropic considerations further expand the allowable wall spacing (see Fig. S3 of the supplemental materials).

By focusing on athermal particles, the model system introduced here demonstrates how a substrate prepatterned with sparse, enthalpic template features can disrupt the otherwise entropically favorable packing of the hcp lattice in favor of a periodic structure with square symmetry. It would be interesting to use FMT as a tool of inverse design to discover which template features and geometries would be optimal—within constraints specified by, e.g., a particular substrate preparation method—for obtaining not only square structures but other open 2D lattice packings of interest (kagome, honeycomb, snub square, etc.).

Moreover, to design a directed assembly process for specific material systems of technological interest, other interparticle interactions would also need to be considered. Such interactions might also be tuned, guided by theoretical methods similar to those discussed in [11-14], to complement the template's role in directing assembly into specific structures. For example, soft repulsive interparticle interactions have been shown to significantly reduce the thermodynamic favorability of the hcp packing over other open target structures of interest even on a pattern-free substrate [37]. We are currently exploring how interparticle interactions can affect the pattern amplification performance of graphoepitaxy, and we will report our findings in a separate publication.

**Acknowledgements:**

Funding for this project was provided by the National Aeronautics and Space Administration through NASA Space Technology Research Fellowship #NNX11AN80H and through the Welch Foundation through grant F-1696. This work also made use of NASCENT Engineering Research Center Shared Facilities supported by the National Science Foundation under Cooperative Agreement No. EEC-1160494. Any opinions, findings and conclusions or recommendations expressed in this material are those of the author(s) and do not necessarily reflect those of NASA or the National Science Foundation.

[1] F. Li, D.P. Josephson, and A. Stein, Angew. Chem. Int. Ed. **50**, 360 (2011)
[2] M. Rycenga, C. M. Cobley, J. Zeng, Weiyang Li, C. H. Moran, Q. Zhang, D. Qin, and Y. Xia, Chem. Rev. **111**, 3669 (2011)
[3] G. A. Ozin, K. Hou, B. V. Lotsch, L. Cademartiri, D. P. Puzzo, F. Scotognella, A. Ghadimi, and J. Thomson, Mater. Today **12**, 12(2009)
[4] P.F. Damasceno, M. Engel, and S.C. Glotzer, Science **337**, 453 (2012)
[5] S. Sacanna, W. T. M. Irvine, P. M. Chaikin, and D. J. Pine, Nature **464**, 575 (2010)
[6] L. Filion, M. Hermes, R. Ni, E. C. M. Vermolen, A. Kuijk, C. G. Christova, J. C. P. Stiefelhagen, T. Vissers, A. van Blaaderen, and M. Dijkstra, Phys. Rev. Lett. **107**, 168302 (2011)
[7] Q. Chen, S.C. Bae, and S. Granick, Nature **469**, 381 (2011)
[8] A. Reinhardt, F. Romano, and J. P. K. Doye, Phys. Rev. Lett. **110**, 255503 (2013)
[9] C. Knorowski, S. Burleigh, and A. Travesset, Phys. Rev. Lett. **106**, 215501 (2011)

[10] Y. Zhang, F. Lu, K.G. Yager, D. van der Lelie and O. Gang, Nature Nanotechnology **8**, 865 (2013)
[11] M. C. Rechtsman, F. H. Stillinger, and S. Torquato, Phys. Rev. Lett. **95**, 228301 (2005)
[12] S. Torquato, Soft Matter **5**, 1157 (2009)
[13] A. V. Tkachenko, Phys. Rev. Lett. **106**, 255501 (2011)
[14] A. Jain, J. R. Errington, and T.M. Truskett, Soft Matter **9**, 3866 (2013)
[15] N. V. Dziomkina and G. J. Vancso, Soft Matter **1**, 265 (2005)
[16] K. Mangold, P. Leiderer, and Clemens Bechinger, Phys. Rev. Lett. **90**, 158302 (2003)
[17] A. van Blaaderen, M. Dijkstra, R. van Roij, A. Imhof, M. Kamp, B. W. Kwaadras, T. Vissers, and B. Liu, Eur. Phys. J. Special Topics **222**, 2895 (2013)
[18] A. C. Mitus, H. Weber, and D. Marx, Phys. Rev. E **55**, 6855 (1997).
[19] R.A. Segalman, H. Yokoyama, and E. J. Kramer, Adv. Mater. **13**, 1152 (2001)
[20] R. A. Segalman, A. Hexemer, and E. J. Kramer, Macromolecules **36**, 6831 (2003)
[21] I. Bita, J.K.W. Yang, Y.S. Jung, C.A. Ross, E.L. Thomas, and K.K. Berggren, Science **321**, 939 (2008)
[22] J.Y. Cheng, C.T. Rettner, D.P. Sanders, H-C. Kim, and W.D. Hinsberg, Advanced Materials **20**, 3155 (2008)
[23] E. Han, H. Kang, C.-C. Liu, P.F. Nealey, and P. Gopalan, Adv. Mater. **22**, 4325 (2010)
[24] Q. Tang, and Y. Ma, Soft Matter **6**, 4460 (2010)
[25] J. K. W. Yang, Y. S. Jung, J.-B. Chang, R. A. Mickiewicz, A. Alexander-Katz, C. A. Ross, and K. K. Berggren, Nature Nanotechnology **5**, 256 (2010)
[26] K. C. Daoulas, A. Cavallo, R. Shenhar, and M. Muller, Phys. Rev. Lett. **105**, 108301 (2010)
[27] X. Man, D. Andelman, and H. Orland, Phys. Rev. E **86**, 010801 (2012)
[28] J. Qin, G.S. Khaira, Y. Su, G.P. Garner, M. Miskin, H.M. Jaeger, and J.J. de Pablo, Soft Matter **9**, 11467 (2013)
[29] Y. Rosenfeld, Phys. Rev. Lett. **63**, 980 (1989)
[30] F-Q. You, Y-X. Yu, and G-H. Gao,J. Phys. Chem. B **109**, 3512 (2005).
[31] R. Roth, K. Mecke, and M. Oettel, J. Chem. Phys. **136**, 081101 (2012)
[32] M.P. Sears and L.J.D. Frink, J. Comp. Phys. **190**, 184 (2003)
[33] G. Doge, K. Mecke, J. Moller, D. Stoyan, R. P. Waagepetersen, Int. J. Mod. Phys. C **15**, 129 (2004)
[34] C. Tang, E. M. Lennon, G. H. Fredrickson, E. J. Kramer, C. J. Hawker, Science **322**, 429 (2008)
[35] B. Widom, J . Chem. Phys. **39**, 2808 (1963)
[36] R. Peikert, D. Würtz, M. Monagan, and C. de Groot, in System Modelling and Optimization, edited by P. Kall (Springer Berlin Heidelberg, 1992), Vol. 180, pp. 45-54.
[37] M. C. Rechtsman, F. H. Stillinger, S. Torquato, Phys. Rev. E **73**, 011406 (2006)

# Supplementary Material

## I. Transition Away from Square Order

In Figure 3 of our letter, we show that the density profile peaks for wall spacing beyond 20R (corresponding to a 10x10 lattice) are no longer pronounced. Analysis of this transition can be further supplemented by extracting a correlation length from the density profiles. This can be obtained by analyzing the decay in the height of local maxima for peaks further away from the template features. To quantify this decay, we introduce a new quantity, $x_c$, defined as the location where the peak height decays by a factor of 1/e. We then observe a cross-section of the density profile taken from a middle row of the system (e.g., $\rho\left(\frac{x}{R}, \frac{L}{2R}\right)$) where wall effects are at their weakest. Density maxima from this cross-section are then normalized between 0 and 1 using $\bar{\rho} = \left(\rho(x/R) - \rho_{avg}\right)/\left(\rho(1) - \rho_{avg}\right)$, where $\rho(1)$ is the density against the wall.

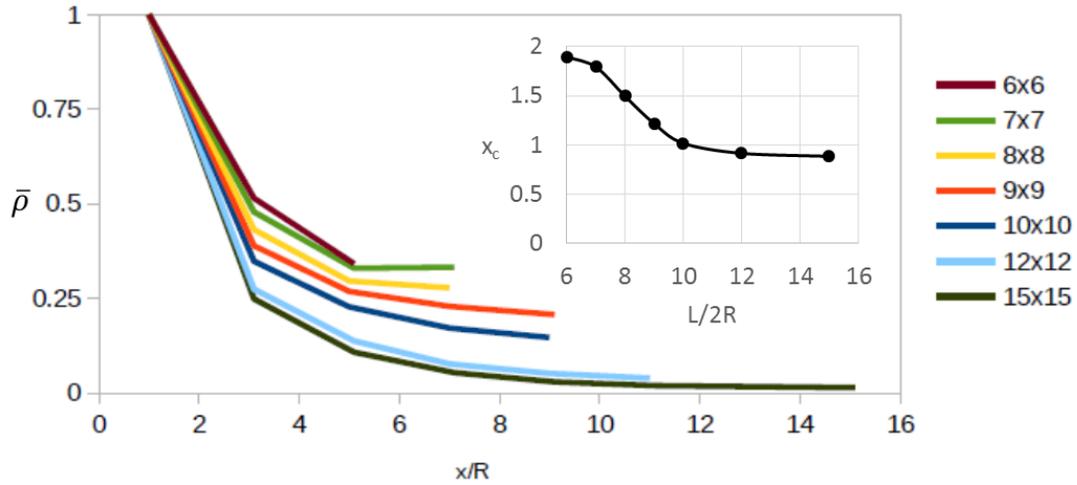

Figure S1: Local maxima of the middle cross-section $\rho\left(\frac{x}{R}, \frac{L}{2R}\right)$ of density profiles for various wall-separations. Here, the density is normalized by $\bar{\rho} = \left(\rho(x/R) - \rho_{avg}\right)/\left(\rho(1) - \rho_{avg}\right)$ and $x_c$ is defined as the value of (x-1)/2R where $\bar{\rho}(x) = 1/e$. The height of the local maxima decreases quickly, with $x_c$ occurring before the second peak for lattices greater than 10x10.

The value of $x_c$ decreases dramatically, approaching the first non-contact peak in the profile by the 10x10 lattice. This pronounced decline in the propagation of wall effects is consistent with the observation that the transition away from square order occurs around the 20R wall separation and, thus, that the corresponding sparser profiles in Figure 3 have such ill-defined peaks. This result is further clarified by grand canonical Monte Carlo simulations, shown in Figure S2, which show a clear difference between the structures formed under the 20R and 24R wall separations.

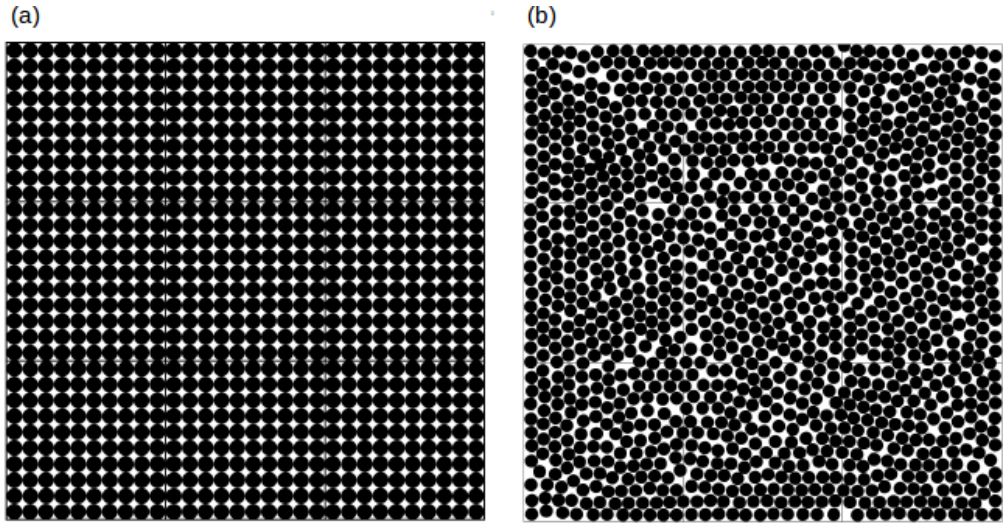

Figure S2: Grand canonical Monte Carlo snapshots of hard sphere monolayers exposed to perpendicular walls separated by (a) 20R and (b) 24R.

## II. Fractional Expansion of Template Spacing

In our Letter, we have shown that a sparse square template promotes the assembly of hard spheres into a square lattice. Our discussion focused on templates composed of four perpendicular, thin barriers separated by a distance commensurate with the square lattice. However, it is interesting to note that the barrier separation need not be an exact integer multiple of the square lattice periodicity (i.e., the particle diameter). The imposed field may be expanded by fractions of a particle radius, and still preserve square order. This is not entirely surprising, as for many lattices, there is a minimal required expansion before a new particle can physically fit within the system. In addition to physically having room for a new particle, it must be thermodynamically favorable to restructure into a more tightly packed lattice. As illustrated in Figure S1, we have found in most cases that template barriers may be spaced at a commensurate length plus an additional ~0.5R in order to preserve the square lattice.

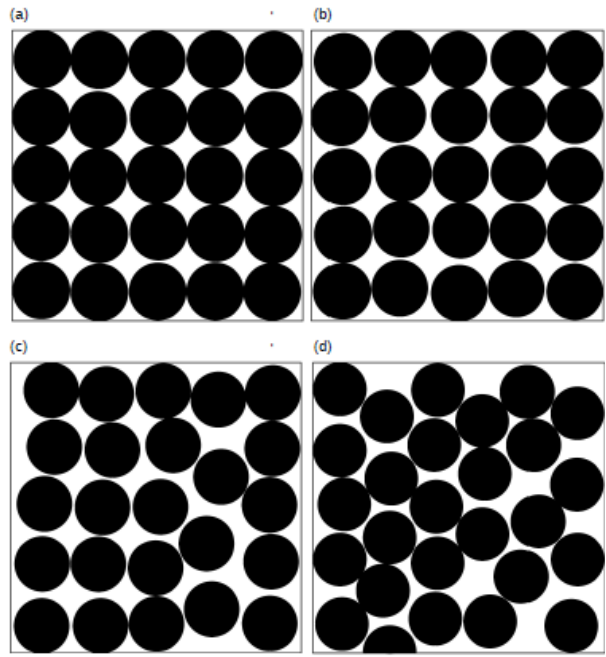

Figure S3: Hard sphere monolayers exposed to an external field of varying wall spacing. Distance between walls is (a) 10.2R, (b) 10.4R, (c) 10.6R, and (d) 11.0R where R is the particle radius.

## III. The Direct Pattern

While our work here has primarily focused on the effects of a sparse template, it is useful to see how a direct template behaves under similar conditions. As shown in Figure S2, a direct template is able to facilitate the formation of a square template at a lower chemical potential than the sparse templates (keeping in mind that direct patterning negates the practical "pattern multiplication" benefits of a sparse template). This effect is due to the template fixing each particle in the ideal square lattice coordinate, even at low density. Thus, it becomes easier for each subsequent particle to enter in the desired location, as the existing particles do not require any restructuring. However, as chemical potential increases, the lattice for the direct template is very similar to that of the sparse template shown in Figure 4(c) of the attached letter.

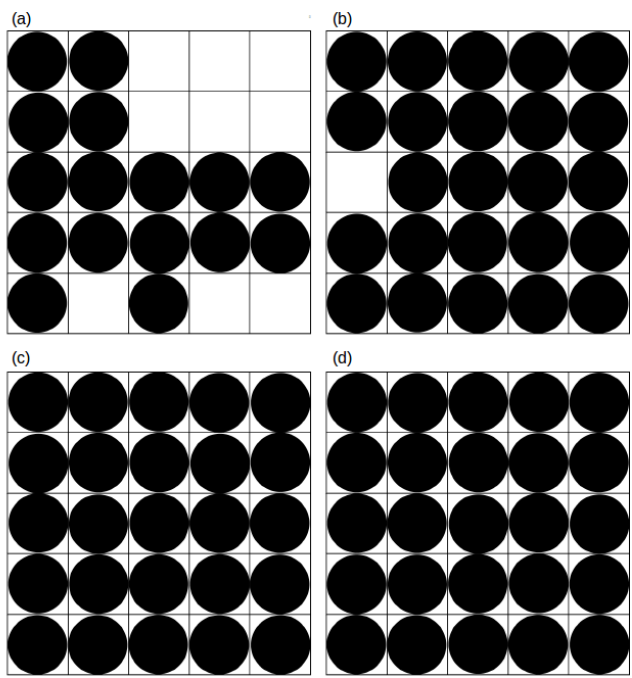

Figure S4: Grand canonical Monte Carlo snapshots of particles exposed to a direct pattern, with perpendicular walls spaced by 2R. Particle deposition is governed by chemical potential, $\beta(\mu-\mu_{hcp})$ = (a) -8.1, (b) -4.1, (c) 0.0, and (d) 3.9. As chemical potential increases, vacant wells are filled.